\documentclass[twocolumn,preprintnumbers, notitlepage,amsmath,superscriptaddress,10pt,aps,pra]{revtex4-1}

\usepackage{graphicx}
\usepackage{color}
\usepackage{hyperref}
\usepackage{subfigure}
\usepackage{graphicx}
\usepackage{type1cm}
\usepackage{eso-pic}
\usepackage{color}
\usepackage{amsmath}
\usepackage{amssymb}
\usepackage[left]{lineno}
\usepackage{blindtext}

\def\blue#1{\textcolor{black}{#1}}
\def\bluee#1{\textcolor{black}{#1}}

\begin{document}



\title{Indistinguishable single photons  with flexible electronic triggering}


\affiliation{Institute for Photonics and Quantum Sciences, SUPA, Heriot-Watt University, Edinburgh EH14 4AS, United Kingdom}
\affiliation{Center for Opto-Electronic Convergence Systems, Korea Institute of Science and Technology, Seoul, Korea}

\author{Adetunmise C. Dada}
 \email{a.c.dada@hw.ac.uk}
\affiliation{Institute for Photonics and Quantum Sciences, SUPA, Heriot-Watt University, Edinburgh EH14 4AS, United Kingdom}
\author{Ted~S.~Santana}
\affiliation{Institute for Photonics and Quantum Sciences, SUPA, Heriot-Watt University, Edinburgh EH14 4AS, United Kingdom}
\author{Ralph~N.~E.~Malein}
\affiliation{Institute for Photonics and Quantum Sciences, SUPA, Heriot-Watt University, Edinburgh EH14 4AS, United Kingdom}
 \author{Antonios Koutroumanis}
\affiliation{Institute for Photonics and Quantum Sciences, SUPA, Heriot-Watt University, Edinburgh EH14 4AS, United Kingdom}

 \author{Yong~Ma} 
 \altaffiliation{Current address: Chongqing Institute of Green and Intelligent Technology, Chinese Academy of Sciences, Chongqing, China 400714}
 \affiliation{Institute for Photonics and Quantum Sciences, SUPA, Heriot-Watt University, Edinburgh EH14 4AS, United Kingdom}
  \altaffiliation{Current address: Chongqing Institute of Green and Intelligent Technology, Chinese Academy of Sciences, Chongqing, China 400714}

 \author{Joanna~M.~Zajac}
\affiliation{Institute for Photonics and Quantum Sciences, SUPA, Heriot-Watt University, Edinburgh EH14 4AS, United Kingdom}
 \author{Ju Y. Lim}
  \altaffiliation{Current address: Korea Photonics Technology Institute, Gwangju 61007, Korea}
\affiliation{Center for Opto-Electronic Convergence Systems, Korea Institute of Science and Technology, Seoul, Korea}
\altaffiliation{Current address: Korea Photonics Technology Institute, Gwangju 61007, Korea}

 \author{Jin D. Song}
\affiliation{Center for Opto-Electronic Convergence Systems, Korea Institute of Science and Technology, Seoul, Korea}
\author{Brian D. Gerardot}
\affiliation{Institute for Photonics and Quantum Sciences, SUPA, Heriot-Watt University, Edinburgh EH14 4AS, United Kingdom}


\begin{abstract}
{\bf A key ingredient for quantum photonic technologies is an on-demand source of indistinguishable single photons. State-of-the-art indistinguishable-single-photon sources typically employ resonant excitation pulses with fixed repetition rates, creating a string of single photons with predetermined arrival times. However, in future applications, an independent electronic signal from a larger quantum circuit or network will trigger the generation of an indistinguishable photon. Further, operating the photon source up to the limit imposed by its lifetime is desirable. Here we report on the application of a true on-demand approach in which we can electronically trigger the precise arrival time of a single photon as well as control the excitation pulse duration, based on resonance fluorescence from a single InAs/GaAs quantum dot. We investigate in detail the effect of finite duration of an excitation $\pi$ pulse on the degree of photon antibunching. Finally, we demonstrate that highly indistinguishable single photons can be generated using this on-demand approach, enabling maximum flexibility for future applications.}\\
\end{abstract}

\maketitle
\twocolumngrid

 \section{Introduction}
Single photons remain prime candidates for realising scalable schemes of quantum communication~\cite{Kimble:2008if} and linear optical quantum computing~\cite{Kok:2007tc,OBrien:2009eu}.  The performance of such schemes rely critically on the 
indistinguishability  of the single photons~\cite{Rohde:2012cn}, in particular for key applications such as quantum repeaters~\cite{Briegel:1998jd} and boson sampling~\cite{spring:izTJeUT2,Broome:2013bv}. Of the various types of single photon sources~\cite{Lounis:2005ex,Shields:2007jt}, semiconductor quantum dot (QD) systems are particularly promising for generating indistinguishable single photons because they offer a robust platform in which a single quantum system can be embedded within semiconductor devices and designed into bright single- and entangled-photon sources. 
 The ideal single-photon source for quantum information processing (QIP) applications is one which generates a pure single photon Fock state {\it on demand}, i.e., in response to an independent trigger signal from a user.
Pulsed resonance fluorescence (RF) has been identified as the optimal way to deterministically generate high-quality photons with minimal dephasing.
However, good quality pulsed RF systems have so far utilised pulsed excitation generated by  lasers with fixed repetition rates ($\sim80\hspace{1pt}$MHz)~\cite{He:2013bq,Muller:2014ir,Huber:2015uj,Ding:2016cy,Loredo:2016vz,Somaschi:2015ut}. 
 While this type of triggering could be said to be \emph{deterministic},  it is not  \emph{on-demand} since a user in this case has limited control over the excitation pulse arrival time and duration.
 
\blue{ Here, we apply a flexible scheme for pulsed RF which triggers the generation of highly indistinguishable single photons such that  true on-demand operation is achieved via real-time electronic control.  
Our system uses a GHz-bandwidth electro-optic modulator (EOM) to modulate the output of a tunable continuous-wave (CW) laser for resonant excitation of a QD emitting at $\sim960$\hspace{1pt}nm.  
  In turn, the EOM is driven by a fast programmable electronic pulse-pattern generator (PPG). Such flexibility will greatly benefit practical applications of single photons in quantum technologies.}
    
 Key performance measures for an on-demand single photon source include: the efficiency, defined as the probability to detect a photon for a given electronic trigger; the purity, defined by the degree of antibunching as quantified by the second-order correlation function at zero delay; the degree of indistinguishability between individual photons as measured, e.g., by the Hong--Ou--Mandel (HOM)-type two-photon interference (TPI) visibility~\cite{PhysRevLett.59.2044};  and crucially, the ability to determine or adjust, on demand, the timing and sequence of trigger pulses.
   
  \begin{figure*}[t!]
\centerline{\includegraphics[width=1\textwidth]{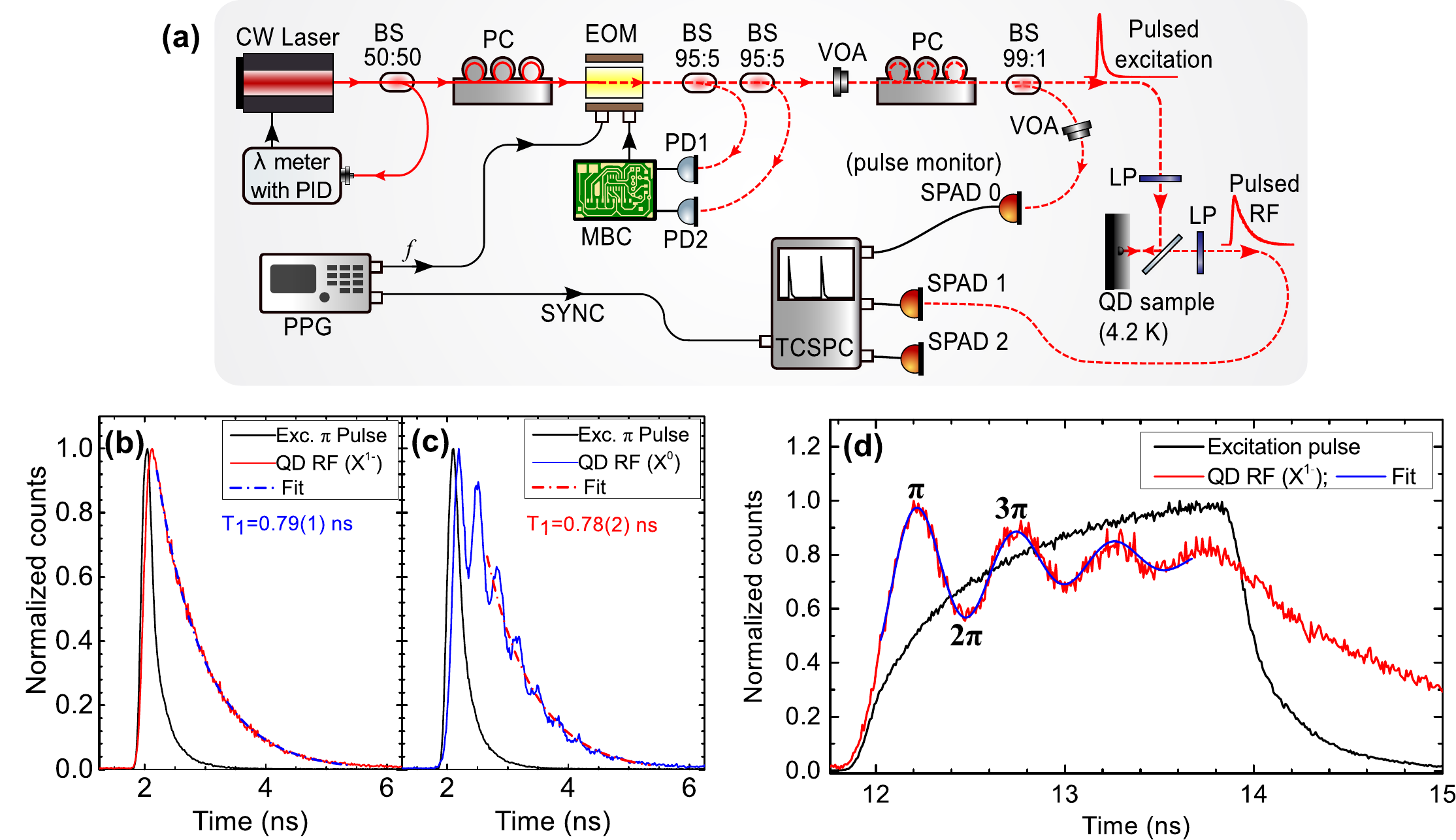}}
\caption{{\bf (a) Flexibly-triggered  generation of resonance fluorescence from a quantum dot.} We modulate the CW laser output using a $20\hspace{1pt}$Gb/s electro-optic modulator (EOM) driven by a pulse-pattern generator (PPG) capable of custom pulse patterns at up to a frequency of $f=3.35$\hspace{1pt}GHz.  A modulator bias controller (MBC) optoelectronic circuit maintains the high extinction ratio of the excitation pulses at $>30$\hspace{1pt}dB \blue{using a dual feedback system for increased dynamic range}.  BS: beam splitter; PC: polarization controller; VOA: variable optical attenuator; LP: linear polarizer; SPAD: single-photon avalanche diode. {\bf (b), (c) Time-resolved QD resonance fluorescence under 100-ps $\pi$-pulse excitation.}  \blue{We overlay pulsed RF on a real-time measurement of the 100-ps excitation pulse (with spectral FWHM $\sim 5.4\hspace{1pt}\mu$eV, see Supplemental Document)  obtained by tapping off some of the power from the EOM output [see (a)]. A fit of a single exponential function to the exciton decay yields lifetimes of $T_1^{X1-}=0.79(1)\hspace{1pt}$ns and  $T_1^{X0}=0.78(2)\hspace{1pt}$ns for  $X^{1-}$ and $X^0$  respectively.  The V-type energy structure of $X^0$ leads to quantum beats between excited states which are directly detected here in the pulsed RF transient decay.  {\bf (d) Direct observation of Rabi oscillations in the charged exciton.} A fit of the theoretical excited state population (see Supplemental Document) to the Rabi oscillations gives a dephasing time $T_2=(2.1\pm0.2\hspace{1pt})T_1$. }  }
\label{fig:eomsetup&pulses}
\end{figure*}
Considerable effort has been made towards realizing on-demand triggering of single photon generation by directly driving a QD electrically. GHz-bandwidth electrical pulses (with pulse width $w > 270\hspace{1pt}$ps) have been used to rapidly modulate the QD emission in resonant or non-resonant excitation~\cite{Prechtel:2012er,Cao:2014ip,Schlehahn:2015dg}. Unfortunately the single photon purity in such hybrid schemes is less than ideal.  

Similar effects have been observed when using an \blue{EOM} to generate optical trigger pulses for a single-photon source (e.g., Ref.~\cite{Matthiesen:2013jq}, $w=500\hspace{1pt}$ps) where  significant overlap between quantum dot RF pulses results in a quasi-CW stream of RF photons. \blue{EOM-generated optical pulses have been used for direct detection of Rabi oscillations in QD excitons ($w=2$\hspace{1pt}ns)~\cite{Schaibley:2013bt}, as well as fast triggering of single-photon generation with a large multi-photon contribution in the emission due to large trigger pulse widths ($w>300$\hspace{1pt}ps)~\cite{Rivoire:2011ft}}. 
    \blue{EOMs have also been applied for triggered photon generation from a QD also using  optical pulses with $w\ge 400 $\hspace{1pt}ps  specifically applied to QD spin manipulation and quantum teleportation~\cite{Gao:2013df}. }
    Efforts have also been made to synchronously modulate QD photoluminescence generated using pulsed \blue{optical} pumping with the goal of waveform shaping and temporal matching~\cite{Rakher:2011ba},  as well as improved single-photon generation \blue{by filtering out multi-photon events and the incoherent portion of the photon wave packets
~\cite{Ates:2013fa}. The uses of EOMs for modulation of single-photon wave packets generated in pulsed mode by non-QD sources have also been demonstrated~\cite{kolchin2008electro,Specht:2009cb}.}

\blue{In all these works, on-demand operation and pure single photon generation of the sources have been undermined by high background counts and the widths of the excitation pulses. We use an EOM to demonstrate narrower optical-excitation-pulse widths and low background counts in the on-demand single-photon emission from a QD, better highlighting the potential of the flexible triggering for high-quality indistinguishable single-photon generation. We also exploit the flexibility of our setup for a detailed experimental study of the effect of finite duration of excitation $\pi$ pulses on the degree of photon antibunching.}


\begin{figure*}
\centerline{\includegraphics[width=1\textwidth]{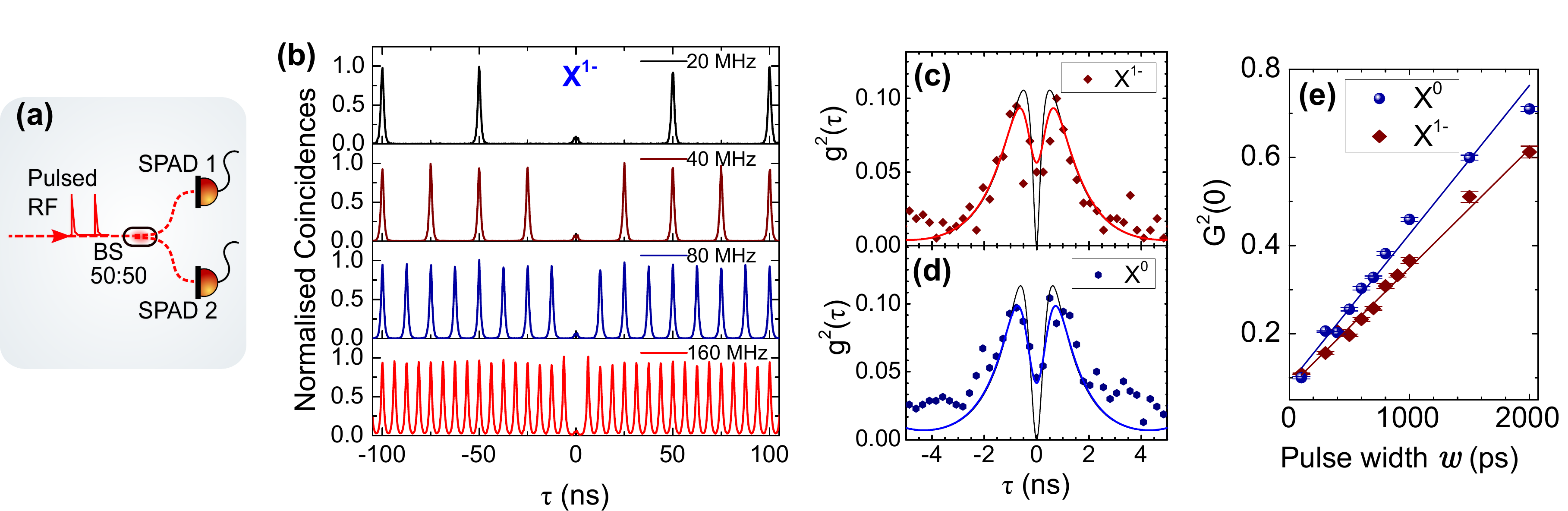}}
\caption{{\bf Pulsed antibunching of on-demand triggered resonance-fluorescence photons}.  (a) Measurement setup (b) Demonstration of flexible triggering of single-photon generation with examples at various frequencies. All measurements have a 180-s integration time.  (c) and (d) show zooms into the time-zero peaks revealing ideal antibunching smeared out by jitter in our detection system (FWHM$\sim 150\hspace{1pt}$ps). The data points represent raw experimental data, while the solid coloured ($g^2(0)\simeq0.05$) and black ($g^2(0)=0.0$) lines respectively represent the results of quantum numerical simulation of the master equation (see Section SII of the Supplemental Document for details) with and without convolution with the instrument response function (IRF) of our detection system (FWHM$\sim 150\hspace{1pt}$ps). The pulsed antibunching is limited by the effect of the finite width of our excitation 100-ps pulses (the limit of our pulse generator) giving $G_{\rm exp}^2(0)\simeq0.1$ and $g_{\rm exp}^2(0)\simeq0.05$. {(e)  {$ G^2(0)$} as a function excitation pulse width under $\pi$-pulse excitation}.  Measurements were performed on both neutral and charged exciton states.  The solid lines are linear fits to the experimental data.}
\label{fig:pulsedG2trend}
\end{figure*}
\begin{figure}[h!]
\centerline{\includegraphics[width=0.5\textwidth]{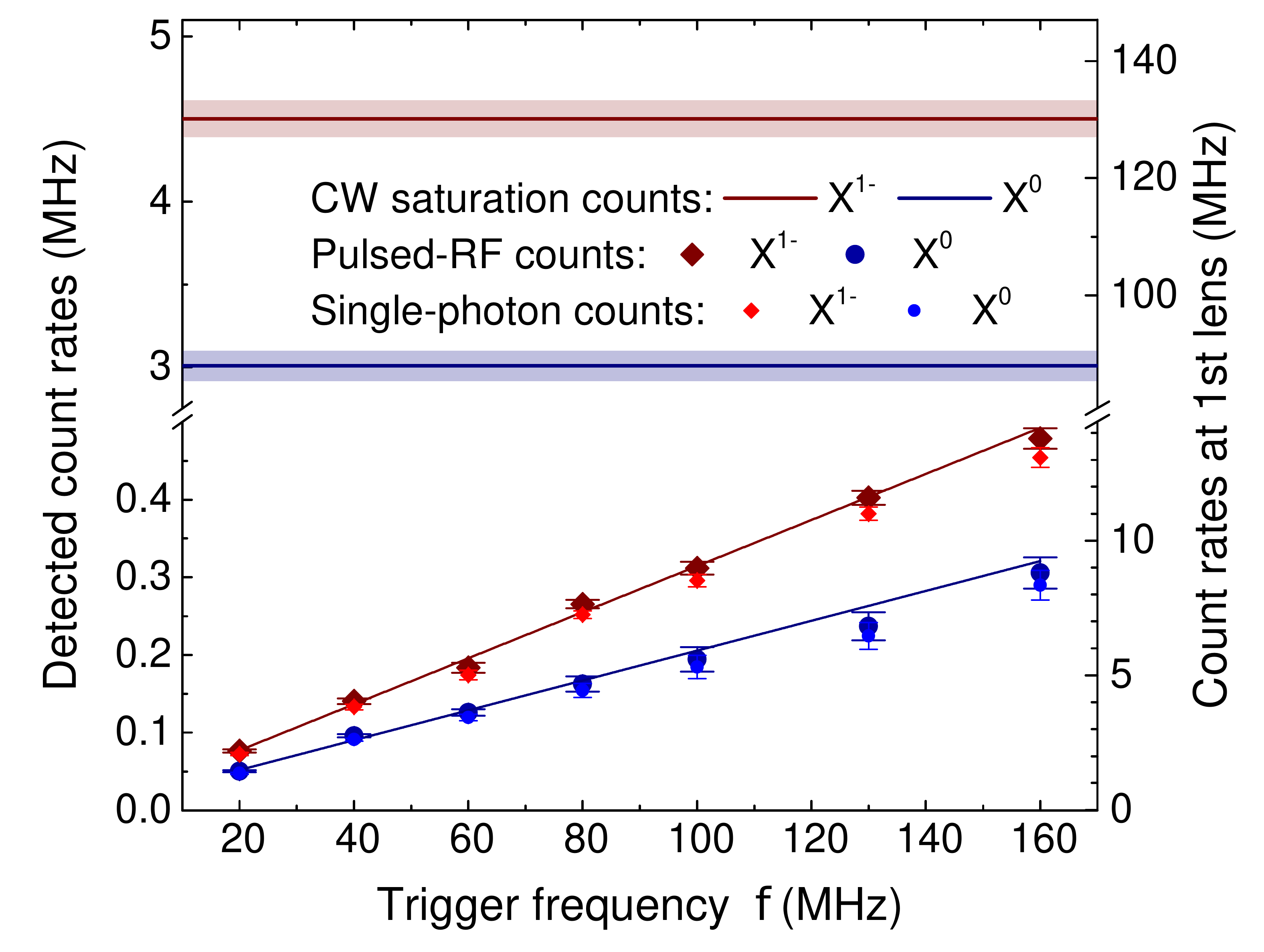}}
\caption{ {\bf Count rates as a function of trigger pulse frequency.}  Raw experimental count rates on  the detector are plotted for both $X^0$ and $X^{1-}$, as well as single photon count rates which are calculated from corresponding multiphoton probabilities [$G^2(0)$]. CW saturation counts are also shown for comparison.}
\label{fig:CountsvFreq}
\end{figure}


\section{Methods}
\noindent
{\bf A. Sample details.}
Our experiments were performed on self-assembled InGaAs quantum dots embedded in a GaAs Schottky diode for deterministic charge-state control. A broadband planar cavity antenna is used to enhance the photon extraction efficiency~\cite{Ma:2014iw}. The QDs are at an antinode of a fifth-order planar cavity on top of a Au layer which functions simultaneously as a cavity mirror and Schottky gate.  Simulations predict a photon extraction efficiency of $ \sim27\%$ into the first objective lens from this device.

\noindent
{\bf B. Resonance fluorescence system.} We perform pulsed RF measurements on both the neutral exciton ($X^0$) and  charged exciton states ($X^{1-}$) of a quantum dot. 
Our setup for triggering single-photon generation on demand is illustrated in Fig.~\ref{fig:eomsetup&pulses}(a). 
For RF, we use a cross-polarization technique in which orthogonally oriented linear polarizers are placed in the excitation and collection arms of a confocal microscope to suppress resonant-excitation-laser photons in the collected light~\cite{Matthiesen:2013jq}, with extinction ratios of more than $10^7$ in CW operation.  

\noindent
{\bf C. Pulsed trigger generation.} 
 We generate our optical excitation trigger signals using a programmable pulse-pattern generator (PPG)  which produces electronic pulses with widths of  down to $w=100\hspace{1pt}$ps at up to  3.35\hspace{1pt}GHz (period $T\simeq300\hspace{1pt}$ps). Notably, much faster pulses ($w<30$\hspace{1pt}ps) can be achieved in the future with commercially available electronic pulse generators. The PPG drives a 20-Gb/s EOM  which in turn modulates the output of a resonant CW laser to obtain optical pulses practically identical to the driving electronic pulses with typical extinction ratios in excess of {30\hspace{1pt}dB}.  This extinction ratio is actively maintained by a modulator bias controller (MBC) optoelectronic circuit through optical feedback. 
 We are able to vary pulse widths and repetition rates of the trigger pulses with high precision, and also obtain optical pulses with user-defined bit-cycle data patterns. 

\noindent {\bf D. Efficiencies.} 
The efficiency of our microscope and detectors are as follows: coupling of far-field radiation into single-mode fiber: $\sim$ 31.4\%; linear polarizer: 43\%; beam splitter surfaces: $(96\%)^4$; SPAD at $\lambda\sim950\hspace{1pt}$nm: $\sim30\%$. The combination gives $\sim 3.5\%$.  The combination gives $\sim3.5\%$. The measured total efficiency of detecting a single photon per trigger pulse is $\sim0.36\%$. Based on this, we determine the photon extraction efficiency from the sample into the first lens to be $\sim10.4\%$.


\section{Results}
Fig.~\ref{fig:eomsetup&pulses}(a) illustrates our basic excitation and measurement setup. 
\blue{In Fig.~\ref{fig:eomsetup&pulses}(b) and (c), we respectively show time-resolved RF from the charged exciton state $X^{1-}$ and $X^{0}$ following excitation with a $\pi$ pulse ($w=100\hspace{1pt}$ps), which gives exciton lifetimes of $T_1^{X1-}=0.79\pm0.01\hspace{1pt}$ns and  $T_1^{X0}=0.78\pm0.02\hspace{1pt}$ns .  The $V$-type energy structure of $X^0$ leads to quantum beats between excited states (e.g., see \cite{Flissikowski:2001hx}) which are directly detected here in the pulsed RF transient decay.  The beat frequency corresponds to the fine-structure splitting (due  to electron-hole exchange interaction) of $\delta_0=3.3$\hspace{1pt}GHz for this QD.}
 In Fig.~\ref{fig:eomsetup&pulses}(d), we demonstrate direct measurement of \blue{$X^{1-}$} Rabi oscillations using 2-ns pulses from which we extract a dephasing time of $T_2=1.66\pm0.18\hspace{1pt}$ns, using the lifetime of $T_1=0.79\pm0.01\hspace{1pt}$ns obtained from the measured \blue{$X^{1-}$} decay with 100-ps pulses. This is consistent with the case of no pure dephasing where $T_2=2T_1$, confirming the absence of excitation induced dephasing effects~\cite{Kyoseva:2005ex,Ramsay:2010gn}.  
    The first  peak in the RF counts corresponds to a pulse area of $\pi$. 
  We confirm the $\pi$-pulse area/power  both using  the direct measurement and by conventional methods (e.g., as used in Ref.~\cite{He:2013bq}). 

\begin{figure*}[t!]
\centerline{\includegraphics[width=1\textwidth]{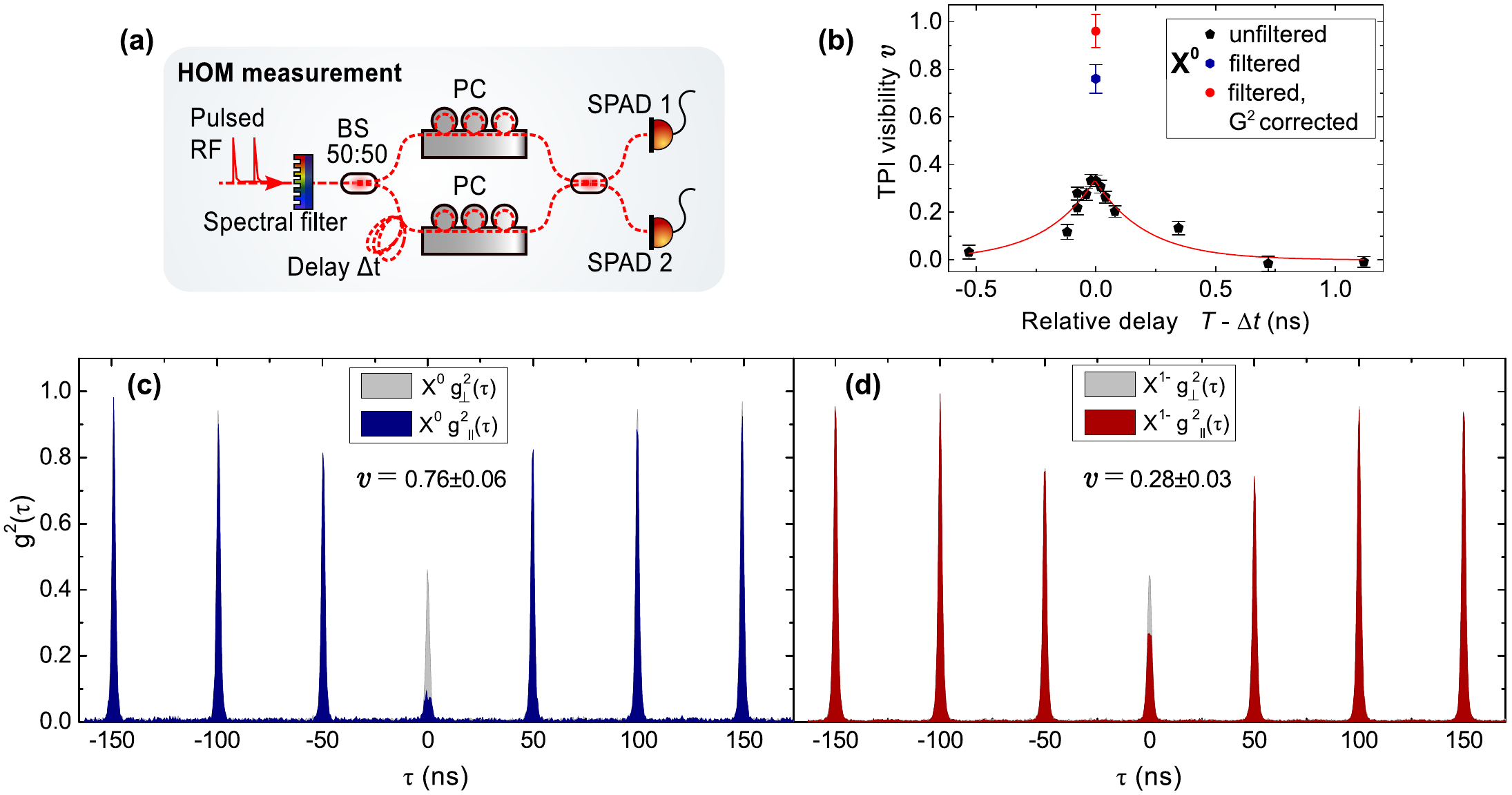}}
\caption{ {\bf Demonstration of indistinguishablility of single photons triggered on demand.}   (a) Hong-Ou-Mandel (HOM)-type two-photon interference (TPI)  results. The flexibility of our approach allows us to set the pulse period to match the delay in our HOM setup ($\Delta t=49.7\hspace{1pt}$ns). (b) {\bf   TPI visibility versus period.} The measurements were perfomed on a neutral exciton line for $ X^{0}$ using various pulse periods and hence delays between interfering photons with $\pi$-pulse excitation. {\bf TPI autocorrelation at zero relative delay.} (c) shows results for $X^{0}$ photons and (d) for the charged exciton ($X^{1-})$ both at $B_{\rm ext}=0\hspace{1pt}$T. Measurements were performed using $100$-ps-wide excitation pulses. The measurements plotted in grey are with orthogonal polarizations of interfering photons.  (c) and (d) are measured with most of the phonon band filtered out using a grating-based spectral filter. The $X^0$ photons show TPI visibiliities of $v=0.76\pm0.06$ as raw experimental data  and $v=0.96\pm0.07$ when corrected \emph{only} for multiphoton emission \blue{($G^2$ corrected)}.}
\label{fig:pulsedHOM}
\end{figure*}


{\bf Antibunching and efficiency.} 
\label{sec:abucngh}
For our main autocorrelation measurements, we
 excite the quantum dot with 100-ps $\pi$ pulses. Fig.~\ref{fig:pulsedG2trend}(a) shows a schematic of the Hanbury Brown and Twiss (HBT)-type set-up used in our antibunching measurements.  In what follows,   while we will use $g^2(\tau)$ to represent the autocorrelation function of the continuous time delay $\tau$,  $G^2(\tau_n)$  denotes the pulsed-mode autocorrelation function of the discretised time delay $\tau_n=nT$  obtained by integrating the $n^{\rm th}$ pulse in $g^2(\tau)$, where $T=1/f$ is the pulse period. In Fig.~\ref{fig:pulsedG2trend}(b), we demonstrate antibunching at various trigger frequencies as seen in the intensity-correlation histograms for the RF emission from the QD under pulsed  excitation. 
  Pulsed second-order correlation at zero delay $G^2(0)$ are calculated by integrating photon counts in the zero-time-delay peak and dividing by the average of the adjacent peaks over a range of  $\sim650\hspace{1pt}$ns around the time-zero peak, with standard deviations obtained from propagated Poissonian counting statistics of the raw counts. With 100-ps $\pi$ pulses, we obtain raw experimental values $G^2(0)\sim 0.1$, and $g^2(0)\sim0.05$, as shown in Figs~\ref{fig:pulsedG2trend} (c-e). An increase in pulse widths leads to worse pulsed antibunching $G^2(0)$ [Fig.~\ref{fig:pulsedG2trend}(e)], while $g^2(0)$ values are unaffected.
 
 We demonstrate the flexibility of the system and how it may be exploited to, e.g., maximize single-photon rates by performing autocorrelation measurements at  varying repetition rates  of $20$\hspace{1pt}MHz to $160$\hspace{1pt}MHz and detect up to  $\sim0.45$\hspace{1pt}MHz of single photon counts (see Fig.~\ref{fig:CountsvFreq}). Also shown are the detected counts rates at saturation in CW mode for each charge state.  Peak coindicence counts of up to $5.5\hspace{1pt}$K are measured at $160\hspace{1pt}$MHz with a 256-ps time bin size in 180-s acquisitions.  Beyond  $\sim160\hspace{1pt}$MHz, the pulses in the RF autocorrelation function begin to overlap. This limit is imposed by the exciton lifetime.  Single-photon count rates are obtained using the emission  probability  of more than one photon in a pulse, as obtained from the corresponding values of $G^2(0)$.  
From the count rates, we calculate the overall efficiency, i.e. probability to detect a pure single photon state per  trigger $\pi$-pulse to be  $0.36\pm0.01\%$.  
Based on the combined efficiency of the collection optics and detectors ($\sim 3.5\%$, see Methods), we determine \bluee{an extraction} efficiency of $10.4\pm0.7\%$ into the first lens \bluee{while the simulated extraction efficiency for our sample is $\sim27\%$ for a 0.68-NA objective lens (as used in our experiment)~\cite{Ma:2014iw}}. 

To reveal the nature of the non-ideal raw antibunching measured in our pulsed experiments and verify the true quality of our single-photon source, we perform high timing resolution (jitter $\sim 150\hspace{1pt}$ps)  measurements of the intensity autocorrelation.  Figs~\ref{fig:pulsedG2trend} (c) and (d)  show zooms into the small time-zero peaks for $X^{1-}$ and $X^{0}$, which both reveal characteristic central dips. At zero delay,  we see clear antibunching within the small peak with a vanishing raw multiphoton probability of  $g^2(0)=0.05$.  
To provide further insight, we use numerical simulations of the master equation for both $X^0$ and $X^{1-}$ (at a magnetic field of $B_{\rm ext}=0$), as a $V$-type atomic system and a two-level system, respectively (see Section SII of the Supplemental Document for details). The small peaks surrounding $\tau=0$ also manifest in the simulation results, in good agreement with  the experimental data as shown in Figs~\ref{fig:pulsedG2trend} (c) and (d). The underlying mechanism of this non-ideality is a small probability to re-excite the system (following a first photon emission event) within the pulse duration. The re-excitation probability increases with pulse width, as confirmed in the simulation and experimentally [see Fig.~\ref{fig:pulsedG2trend}(e)].  Importantly, the $g^2(0)$ is always zero at the middle of the time-zero peak, indicating that arbitrarily low $g^2(0)$ values can be achieved with shorter excitation pulses.  Taking the IRF of our detection system into account, we estimate perfect antibunching ($g^2(0)=0.0$).  
 We conclude that, although more than one photon may be emitted during the 100-ps-long excitation pulse with a small probability, these photons are not emitted at the same time.

{\bf Pulsed two-photon interference.}
For TPI measurements,  we send the QD photons into a HOM-type setup [see Fig.~\ref{fig:pulsedHOM}(a)]  which consists of an unbalanced Mach-Zender (MZ) interferometer with delay of $\Delta t = 49.70$\hspace{1pt}ns  and polarization control in each arm  to enable measurements with parallel ($\parallel$) and orthogonal ($\perp$) polarizations of interfering photons.  
 The beamsplitters in the MZ setup have nearly perfect 50:50 splitting ratios. 
We filter out the zero-phonon line from the most of the phonon sideband using a grating-based spectral filter \blue{(bandwidth $\Delta f = 12\hspace{1pt} $GHz and efficiency $\eta_f=22\%$)}.   
 Due to the flexibility of trigger pulse generation, we are able to precisely match the repetition period of the pulses  
  to $\Delta t$   (see Fig.~\ref{fig:pulsedHOM}(b)) to obtain pulsed autocorrelation at a relative delay $T-\Delta t =0$, shown in Figs \ref{fig:pulsedHOM}(c) and (d). The TPI visibilitiy is defined as    $v=[G^2_{\rm \perp}(0)-G^2_{\rm  \parallel}(0)]/G^2_{\rm \perp}(0)$. For the $X^0$ and $X^{1-}$, we measure raw visibilities of  $v=0.76\pm0.06$ and $0.28\pm0.03$ respectively.  The raw indistinguishability of the $X^{0}$  photons is limited primarily by the multiphoton probability of $G^2(0)=0.10\pm0.01$, which is in turn limited by the excitation pulse width as described above. 
When this is corrected for (by using ${G'}^2_{\rm \parallel}(0)=G^2_\parallel (0)- G^2(0)$), we obtain a TPI visibilities of $v=0.96\pm0.06$ and $0.47\pm0.03$  
respectively for $X^0$ and $X^{1-}$ without accounting for any other experimental imperfections. The reduced visibility of the $X^{1-}$  ($B_{\rm ext}=0\hspace{1pt}$T) is understood to be due to  detuned Raman-scattered photons which are distinguishable from both the elastic and incoherent components of the resonance fluorescence due to nuclear spin fluctuations \blue{(further details are provided in Ref.~\cite{Malein:2015ur}).  We stress that the Raman-scattered photons result in a total linewidth of less than $1$\hspace{1pt}GHz  which is not filtered out by the 12-GHz-bandwidth spectral filter.}

\begin{figure}
\centerline{\includegraphics[width=0.43\textwidth]{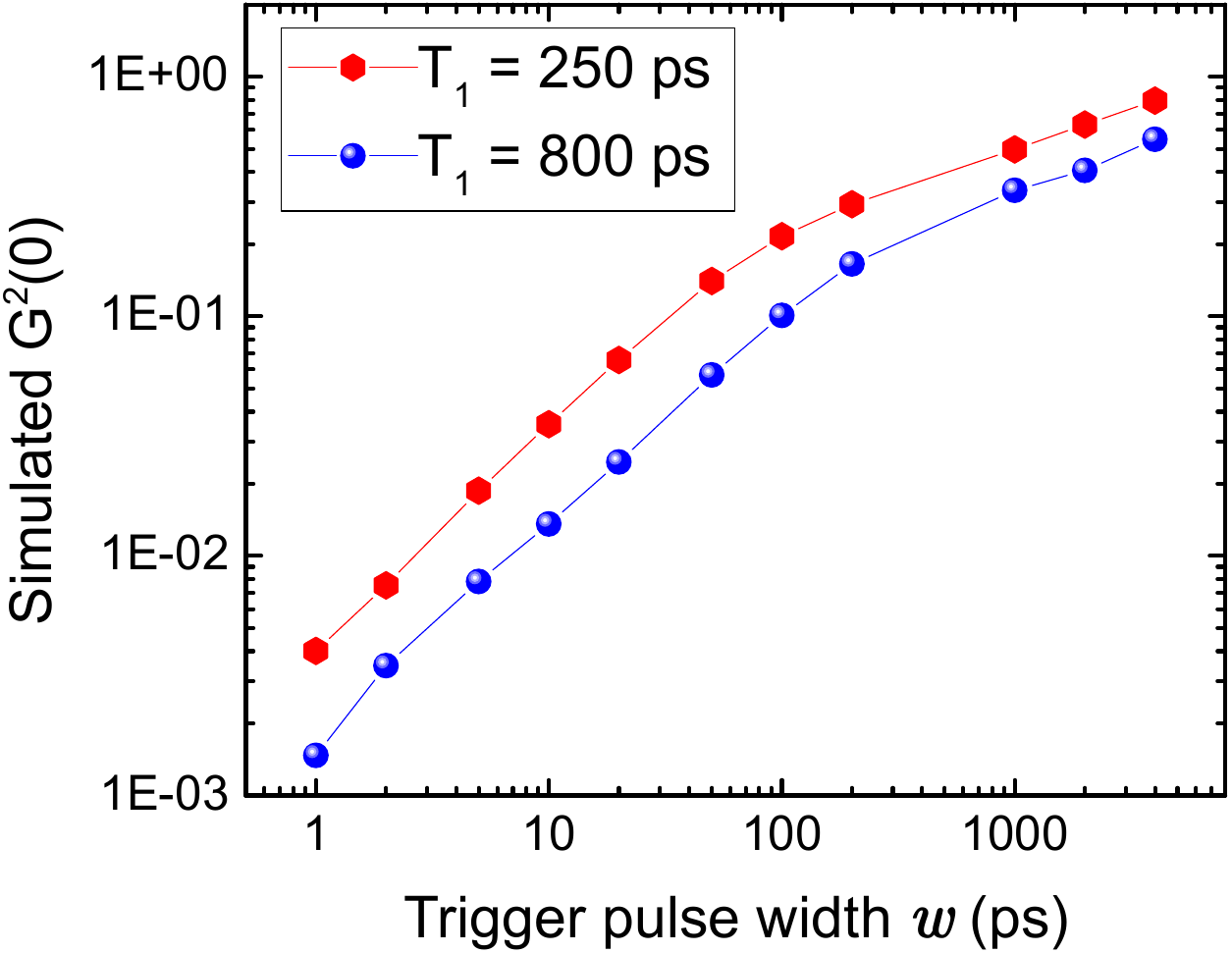}}
\caption{ {\bf Simulated $G^2(0)$ as a function of excitation pulse width under $0.81\pi$-pulse excitation}. Simulation of a two-level system with lifetimes $T_1=800\hspace{1pt}$ps  and $250\hspace{1pt}$ps using a Gaussian (temporal) $0.81\pi$-excitation pulse profiles with varying widths. We use $0.81\pi$ for the simulated Gaussian pulses because with a 100-ps width, they give the same $G^2(0)$ as the asymmetric 100-ps $\pi$ pulses used in the experiment.}
\label{fig:simulationg2}
\end{figure}

\section{Discussion}

For on-demand single photon sources to underpin scalable and efficient linear-optical quantum computing and networking, stringent criteria must be satisfied~\cite{Varnava:2008bd,Jennewein:2011es}. Our experimental results provide insight into the prospect of realizing the $G^2(0)$ requirements using resonance-fluorescence-generated single photons. A crucial result is the effect of the pulse width relative to $T_1$ on $G^2(0)$. Typically, Purcell enhancement is considered desirable to reduce the impact of dephasing mechanisms~\cite{Santori:2002gg,Laurent:2005cb,Gazzano:2013cq}  and enable increased clock rates.  However, in pulsed RF a faster $T_1$ also increases the probability for re-excitation given a certain excitation pulse width. We illustrate this trade-off  using a numerical simulation for $G^2(0)$ as a function of pulse width (Gaussian profile) for $T_1 = 250$\hspace{1pt}ps and 800\hspace{1pt}ps (Fig.~\ref{fig:simulationg2}). We see that in both cases vanishing $G^2(0)$ can be obtained for ultra-short pulse widths, but practically the minimization of $G^2(0)$ is best achieved with larger $T_1$ values.  This is important  for prospective applications (such as linear-optical QIP) of single photons generated using pulsed resonance fluorescence.

We have demonstrated flexible electronic triggering of on-demand single indistinguishable photons. This system offers several intriguing advantages for future applications. Whereas ultra-short excitation pulses lead to excitation induced dephasing (EID)~\cite{Ramsay:2010gn}, coherent control with longer pulse durations is expected to minimize EID~\cite{Forstner:2003ku,Machnikowski:2004id}. Hence, for some coherent control and read-out schemes the flexibility of electronically tunable pulse durations is likely to be attractive. 
\blue{Another advantage of this approach is the possibility to specifically tailor the pump pulses for quantum control processes such as stimulated Raman adiabatic passage~\cite{RevModPhys.70.1003} in quantum dots exhibiting spin-Lambda systems~\cite{PhysRevLett.99.097401,brunner2009coherent}}. 
Finally, we note that the flexible technique presented here enables an excitation repetition rate up to the limit of that imposed by $T_1$, offering a significant boost in count rates for real applications. While overall system efficiencies need to be improved to realize an ideal single-photon source, recent developments in QIP protocols have made efficiency requirements considerably less stringent \blue{(e.g., in \cite{Varnava:2008bd} efficient linear optical quantum computation is possible with an overall efficiency of 2/3)} even as high quality indistiguishablility, antibunching, and brightness are now simultaneously being achieved (e.g., see~\cite{Somaschi:2015ut,Loredo:2016vz,Ding:2016cy}). 
The approach we demonstrate here is an important step towards combining these key performance features with true on-demand operation.


\noindent
{\bf Funding and Acknowledgements.} The authors would like acknowledge the financial support for this work from the Engineering and Physical Sciences Research Council (EPSRC) (EP/G03673X/1, EP/I023186/1, EP/K015338/1) and the European Research Council (ERC) (307392). B.D.G acknowledges the Royal Society for support via a University Research Fellowship. The KIST authors acknowledge support from KIST's flagship program and GRL.


\vspace{20pt}
\noindent




%
\renewcommand{\thefigure}{S\arabic{figure}}
\renewcommand{\theequation}{S\arabic{equation}}
\renewcommand{\thesection}{S\Roman{section}}
\renewcommand{\thesubsection}{s\Roman{section}}
\renewcommand{\thetable}{S\arabic{table}}

\setcounter{figure}{0}
\setcounter{equation}{0}
\setcounter{section}{0}

\newpage

\onecolumngrid 
{\centering
{\bf \large Supplementary Information for: \\~~~~~~~~~~~~``\mbox{Indistinguishable single photons  with flexible electronic~triggering}''}}
\onecolumngrid


\section{Numerical simulation for pulsed resonance fluorescence}
\begin{figure}[h!]
\centerline{\includegraphics[width=0.35\textwidth]{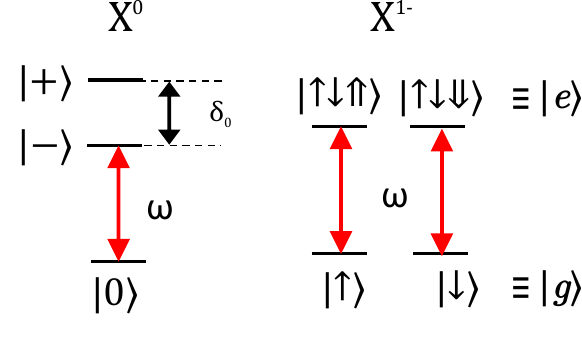}}
\caption{ {\bf Level diagrams for $X^0$ and $X^{1-}$ at an external magnetic field of $ B_{ext}=0$.}  The $X^{1-}$ is modelled as a two-level system (due to the degeneracy of the excited and ground state levels), and the $X^0$ as a V-type three-level atomic system. $\delta_0$ is the electron-hole exchange interaction energy, the red arrows represent the driving field and $\omega$ its frequency. }
\label{fig:leveldiag}
\end{figure}

Our numerical simulation is based on the master equation method~(e.g., see~\cite{puri2001mathematical,breuer2002theory,orszag2007quantum}). We use the well known Hamiltonians describing a driven (1)  two-level  atom  for the $X^{1-}$ and (2) $V$-type three level atom for $X^0$ state of the InGaAs quantum dot (see Fig.~\ref{fig:leveldiag}).

The Hamiltonian for the two-level system is 
\begin{align}
\label{eq:HamiltonianX1-}
\hat{H}_{X1-} = \frac{\Delta}{2} (|e\rangle\langle e| -  |g\rangle\langle g| )+ \frac{\Omega}{2} (|g\rangle\langle e| +  |e\rangle\langle g|), 
\end{align}
while for  $X^0$, we have
\begin{align}
\label{eq:HamiltonianX1-}
\hat{H}_{X0} = &\frac{\Delta}{2} (|-\rangle\langle -| -  |0\rangle\langle 0|) + (\frac{\Delta}{2}+\delta_0)|+\rangle\langle +|  \nonumber+\\ &\frac{\Omega}{2}\left[ \cos\theta\left(|-\rangle\langle 0|+|0\rangle\langle -|\right)+\sin\theta(|+\rangle\langle 0|+|0\rangle\langle +|) \right] ,
\end{align}
under the rotating-wave approximation. $\Delta$  is the detuning between the driving field and the transition from  $|0\rangle$ ($|g\rangle$) to  $|-\rangle$ ($|e\rangle$), $\delta_0$ the electron-hole interaction energy (corresponding to a fine-structure splitting of 13\hspace{1pt}$\mu$eV for the QD reported in Figs 1-4 of the manuscript).  $\Omega$ represents the Rabi frequency,  $\theta$ is a constant determined by  the polarization angle of the (linearly polarized) driving field. 

Spontaneous decay and dephasing  are included in the time evolution of the system via the Lindblad terms of the master equation. The Lindblad operator acting on the density matrix $\rho$ (for a given collapse operator $\hat{C}$) is defined as 
\begin{align}
\label{eq:linblad}
L(\hat{C})\rho = \hat{C}\rho\hat{C}^\dag-\frac{1}{2}( \hat{C}^\dag\hat{C}\rho +\rho\hat{C}^\dag\hat{C})
\end{align}
The master equation can then be written as
\begin{align}
\label{eq:mastereq}
\dot{\rho} = -\frac{i }{\hbar}[\rho,\hat{H}]+\sum_{ij}\Gamma_{ij}L(\sigma_{ij})\rho.
\end{align}
Here, $ \sigma_{ij}=|i\rangle\langle j|$, $\Gamma_{ij}$ is the decay rate from state $|j\rangle$ to $|i\rangle$ and  $\Gamma_{ii}$ represents the pure dephasing rate of state $|i\rangle$, where $i,j = e,g, 0, +, -$. For example, $\Gamma_{eg}=1/T_1$,  $\Gamma_{ee}=\Gamma_{gg}=1/{T_2}-1/{2 T_1}=0$ for $T_2=2T_1$  where $T_1$ is the exciton lifetime, and $1/T_2$ is the total dephasing rate.

Using the Runge-Kutta fourth-order method, we obtain numerical solutions of the master equation and calculate the autocorrelation function from the obtained density matrix elements. 
We simulate pulsed excitation by incorporating the corresponding temporal pulse intensity profile and pattern into the driving field  $\Omega(t)$ as $I(t) \propto \Omega(t)^2$. 

While we use Gaussian temporal pulse profiles to obtain the simulation results in the trend shown in Fig.~5 of the primary manuscript, for the results shown in Fig.~2(c) and (d), we use a $\sim$100-ps-wide lognormal temporal profile  which is obtained by fitting to the intensity profile measured from the EOM output. This closely matches the (asymmetric) intensity profile of our excitation pulses. 

\blue{\section{Modulation setup}}
We use a z-cut electro-optic modulator having a dual bias-port (for coarse- and fine-tuning) with $V_\pi=2$\hspace{1pt}V to cause the phase shift required to change from minimum to maximum transmission. The EOM is driven by a PPG with a 12.5 Gb/s pulse amplifier at the output to ensure full modulation depth.   As mentioned in the main manuscript, the high modulation extinction ratio is achieved and actively maintained  by a modulator bias controller which uses two photodetectors, with different sensitivities. The lower(higher)-sensitivity detector is used in the coarse(fine) tuning of the modulator bias. Two 95:5 beam splitters are used to tap off light from the EOM output for coupling to the respective photo-detectors. 

\vspace{20pt}
\section{Spectral profile of excitation pulses}
\begin{figure}[h!]
\centerline{\includegraphics[width=0.45\textwidth]{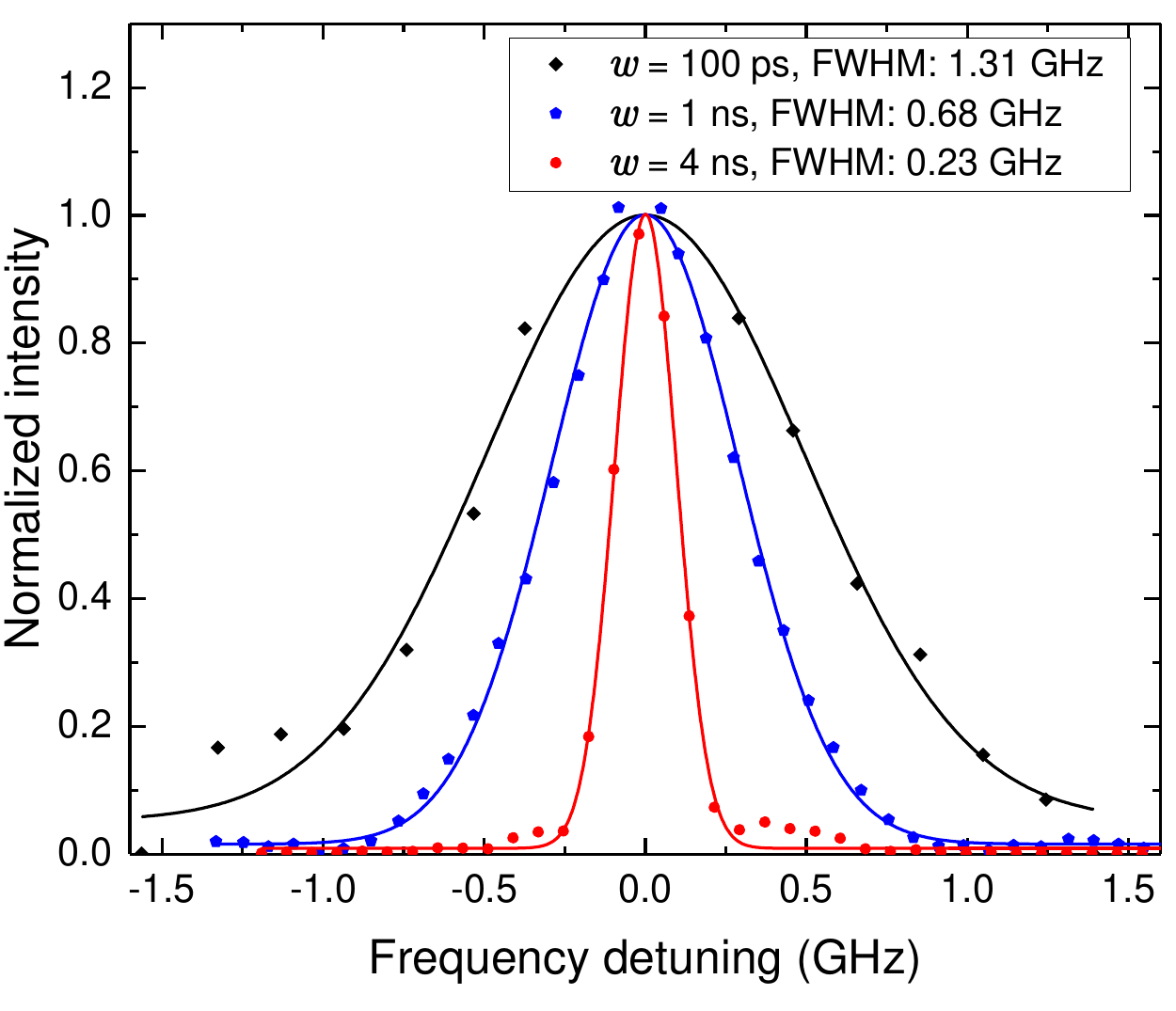}}
\caption{ {\bf High resolution spectrum of excitation pulses at different pulse widths ${w}$.}  The points and lines respectively represent experimental data and fits of Gaussian functions to the data, from which we obtain a full-width-at-half-maximum (FWHM) of $1.3\pm0.1$ GHz ($5.4\pm0.4\hspace{1pt}\mu$eV) for the 100-ps pulses.}
\label{fig:freqdomain}
\end{figure}

We characterise the spectral profile of the 100-ps pulses from the output of the electro-optic modulator (EOM) used in the experiment with a Fabry-Perot scanning interferometer having 27-MHz resolution and a free spectral range of  5.5\hspace{1pt}GHz. There is a broadening of the modulated light in the spectral domain as compared with the continuous-wave (CW) light, with longer pulses exhibiting narrower spectral widths as expected. Fig.~\ref{fig:freqdomain} compares the measured frequency-domain profile of the 100-ps pulses used for the two-photon interference and antibunching measurements reported in the manuscript with those of 1-ns and 4-ns pulses. 

\begin{figure}[ht!]
\centerline{\includegraphics[width=0.45\textwidth]{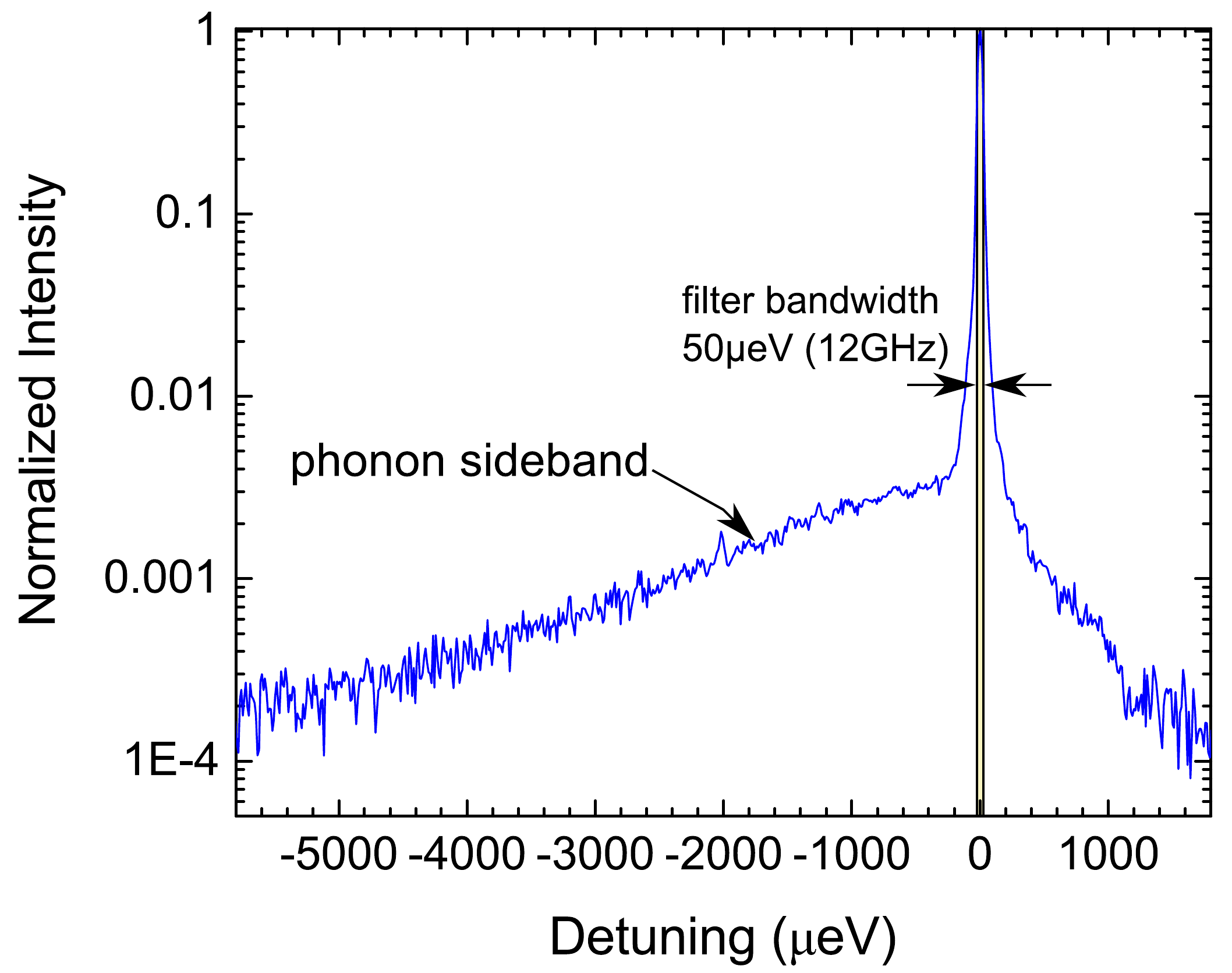}}
\caption{ {\bf High-power spectrum of RF photons under 100-ps pulsed excitation} .  Typical spectral data for a 100-ps pulsed RF from a single quantum dot ($\lambda \sim 960$\hspace{1pt}nm) in our sample with no filtering plotted on log scale to show the phonon sideband  
and showing the bandwidth of the filter used in the Hong-Ou-Mandel-type TPI measurements. This spectral data was  acquired using a 35-$\mu$eV resolution grating spectrometer. The linewidth of the zero-phonon line shown here is limited by spectrometer resolution.  
We note that  $>87$\% of the total signal spectra lies within the pass band of the filter.
}
\label{fig:freqdomainspectro}
\end{figure}

\blue{\section{ RF spectrum/filtering in TPI measurements}}
As mentioned in the main manuscript, the RF photons were spectrally filtered using a 12-GHz-bandwidth grating-based filter for the TPI experiment. We show typical spectral data for a 100-ps pulsed RF from a quantum dot in our sample in Fig.~\ref{fig:freqdomainspectro} to compare the unfiltered RF spectral widths with the filter bandwidth. 
We show the pulsed RF spectrum on a Log scale to highlight the phonon sideband. In this QD the phonon sideband to zero-phonon line ratio is $\sim1:9$.

\vspace{10pt}

 \begin{figure}[h!]
\centerline{\includegraphics[width=0.45\textwidth]{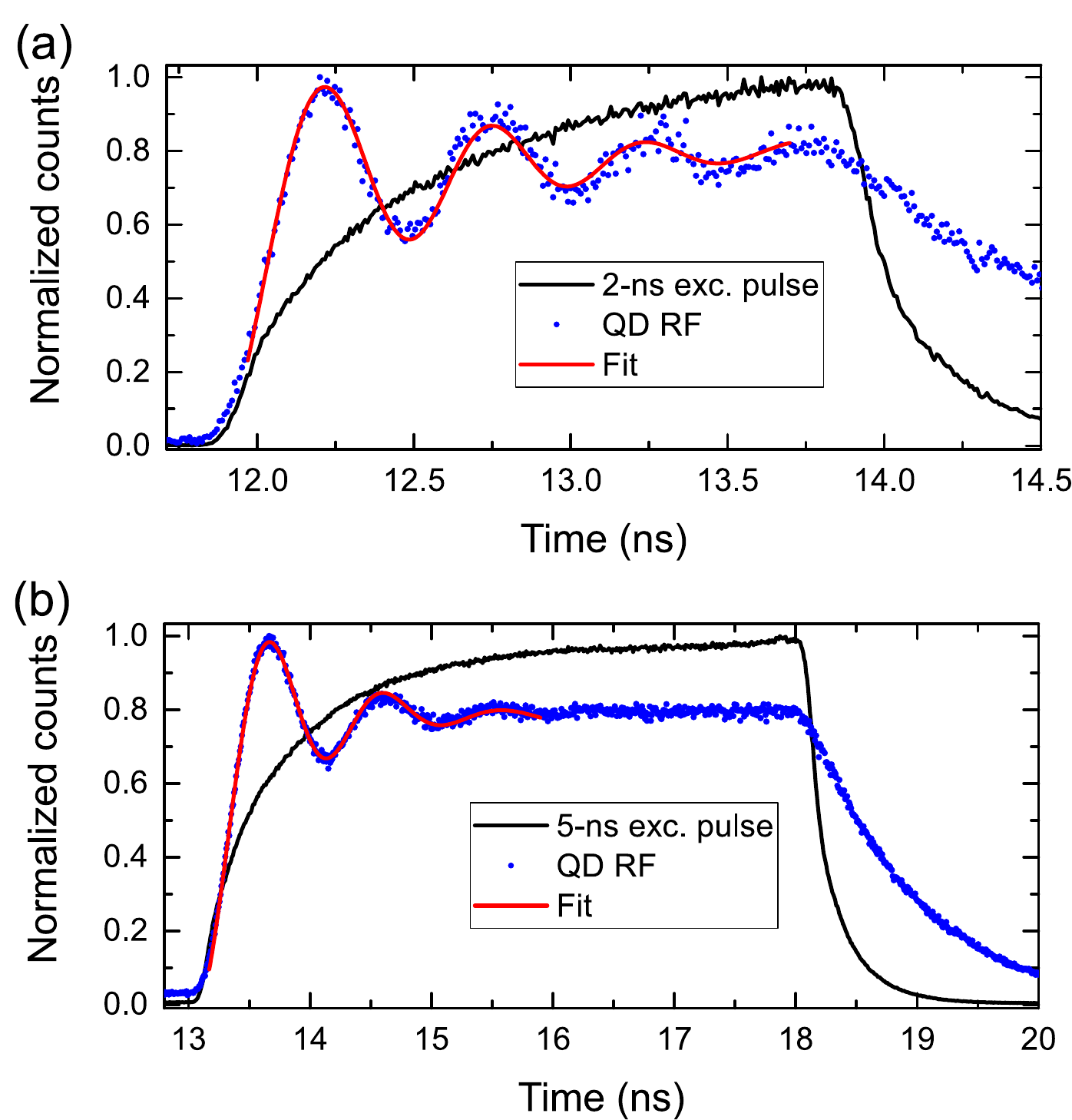}}
\caption{ {\bf $X^{1-}$ Rabi oscillations measurement} using (a) 2-ns  at $\sim188.5\hspace{1pt}$ nw peak power and (b) 5-ns excitation pulses at  $\sim55.5\hspace{1pt}$nw peak power. The red line represents a fit of the theoretical excited state population to the Rabi oscillations when the shape of the excitation pulse is taken into account, i.e., using Eq.~
\ref{eq:rabifreqt} in Eq.~\ref{eq:RabiOscFit} for the fit  gives a dephasing time $T_2 =(2.02\pm0.27)T_1$.}
\label{fig:2ns_RabiOsc3b}
\end{figure}

  \blue{\section{Fitting function for Rabi-oscillations measurement}}
 The function fitted to the Rabi oscillation data describing the probability of being in the excited state is obtained by solving the density-matrix equations for a resonant driving field, including the effects of decoherence due to spontaneous emission and pure dephasing, which is a well known result (e.g., it is the same result used to fit Rabi-oscillation data as reported in Ref.~\cite{Schaibley:2013bt}), i.e., 
\begin{align}
\label{eq:RabiOscFit}
\rho_{22}(t) = &\frac{\Omega^2/2}{\Omega^2+1/{(T_1T_2)}} \nonumber\\
&\times \left\{ 1-\left[  \cos(\xi t) + \tfrac{1/T_1+1/T_2}{2\xi} \sin{(\xi t)}\right] e^{-\tfrac{1}{2}(1/T_1+1/T_2)t}\right\}.
 \end{align}
Here, 
\begin{align}
\label{eq:RabiOscFit_xi}
\xi = \sqrt{\Omega^2-\frac{(1/T_2-1/T_1)^2}{4}},
 \end{align}
$\Omega$ is the Rabi frequency, and $T_1,T_2$ are again the exciton lifetime and the total dephasing time respectively.

\vspace{10pt}

 \begin{figure}[h!]
\centerline{\includegraphics[width=0.4\textwidth]{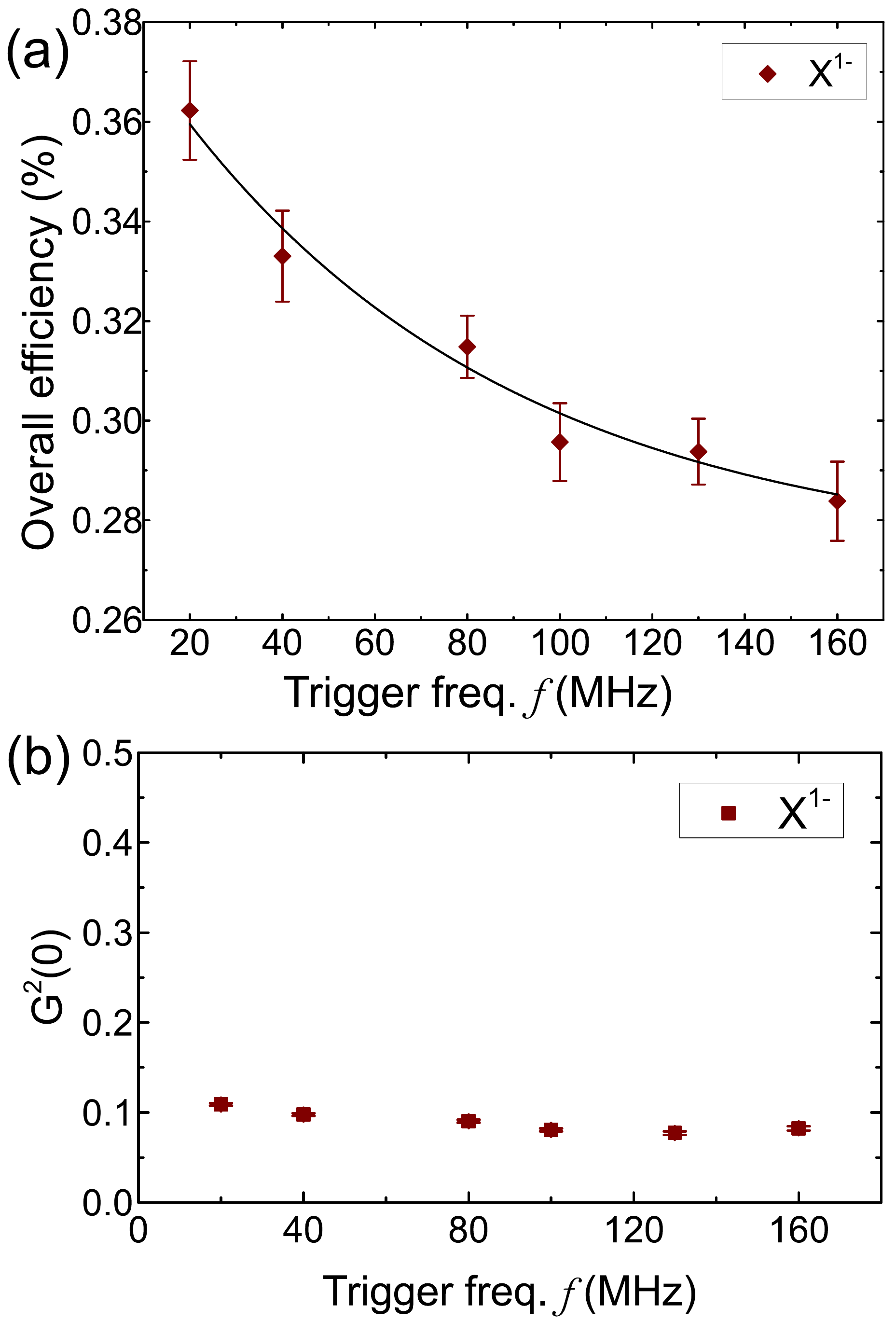}}
\caption{ {\bf (a) Efficiency and (b) $G^2$ trend with varying trigger frequency}  under excitation with 100-ps $\pi$ pulses. An exponential decay was fit to \bluee{the} data in (a). We observe similar values for $G^2(0) (\approx0.1)$ over the range of trigger frequencies $20 $ to $160$\hspace{1pt}MHz. 
}
\label{fig:g2efftrendf}
\end{figure}

 \blue{\section{Temporal profile of excitation pulses for Rabi oscillation measurements}}
The 2-ns optical excitation pulse which were used for the Rabi oscillation measurement shown in Fig. 1 (d) of the main manuscript is a faithful reproduction of the electrical pulse from the PPG, which has a slow rising edge.  
The effect of the increasing power over the duration of the 2-ns pulse is to cause a slight chirp (i.e., variation with time) in the frequency of Rabi oscillation $\Omega$.  
  For the result presented in the main manuscript, this effect is not modelled in the function used to fit the Rabi oscillations because it has negligible consequence for the main result of $T_2=2T_1$.  
  
  Here, we include the effect of the temporal profile of the driving field in the fitting function, by setting $\Omega$ not as a constant but as 
  \begin{equation}
 | \Omega(t)|^2\propto(P_0 + P_1 t + P_2 t^2)
  \label{eq:rabifreqt}
  \end{equation}

 This quadratic function gives an excellent fit to the varying excitation power over the region of the Rabi oscillation fit to the RF from which we obtain the actual parameters $P_0$, $P_1$ and $P_2$ which we then use to define the time variation of $\Omega$. In Fig.~\ref{fig:2ns_RabiOsc3b}, we show examples of fits obtained when we use Eq.~
\ref{eq:rabifreqt} in Eq.~\ref{eq:RabiOscFit} for the fit. This gives $T_2=(2.02\pm0.27)T_1$ in agreement with the case used to fit the data in the main manuscript  in which the excitation pulse is assumed to be flat. For comparison, we also show the measurement obtained using 5-ns pulses which gives a consistent result.

 \blue{\section{Efficiency and pulsed antibunching as a function of trigger frequency}}
We show plots of $G^2(0)$ and overall efficiency as a function of trigger frequency under excitation with 100-ps $\pi$ pulses.  As shown in Fig.~\ref{fig:g2efftrendf}(a), we find that the overall efficiency (i.e., the ratio of single-photon count rate and trigger rate) reduces as a function of trigger frequency with $\pi$-pulse excitation.  We understand this to be mainly due to  an artifact  caused by the dead time of our TCSPC system which is $\le 80$\hspace{1pt}ns. In Fig.~\ref{fig:g2efftrendf}(b),   $G^2(0)\sim0.1$ is fairly constant until $f\sim160$\hspace{1pt}MHz. At this point, we observe an onset of a quasi-CW operation caused by overlap between peaks in $g^2(\tau)$ leading to a background of $g^2(\tau)_{min}\sim 0.03$ (see Fig. 2 (b) of the main manuscript).  The trigger frequency at which this occurs is determined by the lifetime $T_1$.

\vspace{10pt}


\begin{thebibliography}{10}
\newcommand{\enquote}[1]{``#1''}

\bibitem{Kimble:2008if}
H.~J. Kimble, \enquote{{The quantum internet},} Nature \textbf{453}, 1023--1030
  (2008).

\bibitem{Kok:2007tc}
P.~Kok, W.~J. Munro, K.~Nemoto, T.~C. Ralph, J.~P. Dowling, and G.~J. Milburn,
  \enquote{{Linear optical quantum computing with photonic qubits},} Reviews of
  Modern Physics \textbf{79}, 135--174 (2007).

\bibitem{OBrien:2009eu}
J.~L. O{\textquoteright}Brien, A.~Furusawa, and J.~Vu{\v c}kovi{\'c},
  \enquote{{Photonic quantum technologies},} Nature Photonics \textbf{3},
  687--695 (2009).

\bibitem{Rohde:2012cn}
P.~P. Rohde and T.~C. Ralph, \enquote{{Error tolerance of the boson-sampling
  model for linear optics quantum computing},} Physical Review A \textbf{85},
  022332 (2012).

\bibitem{Briegel:1998jd}
H.~J. Briegel, W.~Dur, J.~I. Cirac, and P.~Zoller, \enquote{{Quantum Repeaters:
  The Role of Imperfect Local Operations in Quantum Communication},} Physical
  Review Letters \textbf{81}, 5932--5935 (1998).

\bibitem{spring:izTJeUT2}
J.~B. Spring, B.~J. Metcalf, P.~C. Humphreys, W.~S. Kolthammer, X.-M. Jin,
  M.~Barbieri, A.~Datta, N.~Thomas-Peter, N.~K. Langford, D.~Kundys, J.~C.
  Gates, B.~J. Smith, P.~G.~R. Smith, and I.~A. Walmsley, \enquote{{Boson
  Sampling on a Photonic Chip},} Science \textbf{339}, 798--801 (2013).

\bibitem{Broome:2013bv}
M.~A. Broome, A.~Fedrizzi, S.~Rahimi-Keshari, J.~Dove, S.~Aaronson, T.~C.
  Ralph, and A.~G. White, \enquote{{Photonic Boson Sampling in a Tunable
  Circuit},} Science \textbf{339}, 794--798 (2013).

\bibitem{Lounis:2005ex}
B.~Lounis and M.~Orrit, \enquote{{Single-photon sources},} Reports on Progress
  in Physics \textbf{68}, 1129 (2005).

\bibitem{Shields:2007jt}
A.~J. Shields, \enquote{{Semiconductor quantum light sources},} Nature
  Photonics \textbf{1}, 215--223 (2007).

\bibitem{He:2013bq}
Y.-M. He, Y.~He, Y.-J. Wei, D.~Wu, M.~Atature, C.~Schneider, S.~H{\"o}fling,
  M.~Kamp, C.-Y. Lu, and J.-W. Pan, \enquote{{On-demand semiconductor
  single-photon source with near-unity indistinguishability},} Nature
  Nanotechnology \textbf{8}, 213--217 (2013).

\bibitem{Muller:2014ir}
M.~M{\"u}ller, S.~Bounouar, K.~D. J{\"o}ns, M.~Gl{\"a}ssl, and P.~Michler,
  \enquote{{On-demand generation of indistinguishable polarization-entangled
  photon pairs},} Nature Photonics \textbf{8}, 224--228 (2014).

\bibitem{Huber:2015uj}
T.~Huber, D.~F{\"o}ger, G.~Solomon, and G.~Weihs, \enquote{{Optimal excitation
  conditions for indistinguishable photons from quantum dots},}
  arXiv:1507.07404  (2015).

\bibitem{Ding:2016cy}
X.~Ding, Y.~He, Z.~C. Duan, N.~Gregersen, M.~C. Chen, S.~Unsleber, S.~Maier,
  C.~Schneider, M.~Kamp, S.~H{\"o}fling, C.-Y. Lu, and J.-W. Pan,
  \enquote{{On-Demand Single Photons with High Extraction Efficiency and
  Near-Unity Indistinguishability from a Resonantly Driven Quantum Dot in a
  Micropillar},} Physical Review Letters \textbf{116}, 020401 (2016).

\bibitem{Loredo:2016vz}
J.~C. Loredo, N.~A. Zakaria, N.~Somaschi, C.~Anton, L.~D. Santis, V.~Giesz,
  T.~Grange, M.~A. Broome, O.~Gazzano, G.~Coppola, I.~Sagnes, A.~Lemaitre,
  A.~Auffeves, P.~Senellart, M.~P. Almeida, and A.~G. White, \enquote{{Scalable
  performance in solid-state single-photon sources},} arXiv:1601.00654  (2016).

\bibitem{Somaschi:2015ut}
N.~Somaschi, V.~Giesz, L.~De~Santis, J.~C. Loredo, M.~P. Almeida, G.~Hornecker,
  S.~L. Portalupi, T.~Grange, C.~Anton, J.~Demory, C.~Gomez, I.~Sagnes, N.~D.
  Lanzillotti-Kimura, A.~Lema{\^\i}tre, A.~Auff{\`e}ves, A.~G. White, L.~Lanco,
  and P.~Senellart, \enquote{{Near-optimal single-photon sources in the solid
  state},} Nature Photonics Advance Online Publication
  doi:10.1038/nphoton.2016.23 (2016) .

\bibitem{PhysRevLett.59.2044}
C.~K. Hong, Z.~Y. Ou, and L.~Mandel, \enquote{{Measurement of subpicosecond
  time intervals between two photons by interference},} Physical Review Letters
  \textbf{59}, 2044--2046 (1987).

\bibitem{Prechtel:2012er}
J.~H. Prechtel, P.~A. Dalgarno, R.~H. Hadfield, J.~McFarlane, A.~Badolato,
  P.~M. Petroff, and R.~J. Warburton, \enquote{{Fast electro-optics of a single
  self-assembled quantum dot in a charge-tunable device},} Journal of Applied
  Physics \textbf{111}, 043112 (2012).

\bibitem{Cao:2014ip}
Y.~Cao, A.~J. Bennett, D.~J.~P. Ellis, I.~Farrer, D.~A. Ritchie, and A.~J.
  Shields, \enquote{{Ultrafast electrical control of a resonantly driven single
  photon source},} Applied Physics Letters \textbf{105}, 051112 (2014).

\bibitem{Schlehahn:2015dg}
A.~Schlehahn, M.~Gaafar, M.~Vaupel, M.~Gschrey, P.~Schnauber, J.~H. Schulze,
  S.~Rodt, A.~Strittmatter, W.~Stolz, A.~Rahimi-Iman, T.~Heindel, M.~Koch, and
  S.~Reitzenstein, \enquote{{Single-photon emission at a rate of 143?MHz from a
  deterministic quantum-dot microlens triggered by a mode-locked
  vertical-external-cavity surface-emitting laser},} Applied Physics Letters
  \textbf{107}, 041105 (2015).

\bibitem{Matthiesen:2013jq}
C.~Matthiesen, M.~Geller, C.~H.~H. Schulte, C.~Le~Gall, J.~Hansom, Z.~Li,
  M.~Hugues, E.~Clarke, and M.~Atature, \enquote{{Phase-locked
  indistinguishable photons with synthesized waveforms from a solid-state
  source},} Nature Communications \textbf{4}, 1600 (2013).

\bibitem{Schaibley:2013bt}
J.~R. Schaibley, A.~P. Burgers, G.~A. McCracken, D.~G. Steel, A.~S. Bracker,
  D.~Gammon, and L.~J. Sham, \enquote{{Direct detection of time-resolved Rabi
  oscillations in a single quantum dot via resonance fluorescence},} Physical
  Review B \textbf{87}, 115311--5 (2013).

\bibitem{Rivoire:2011ft}
K.~Rivoire, S.~Buckley, A.~Majumdar, H.~Kim, P.~Petroff, and J.~Vu{\v
  c}kovi{\'c}, \enquote{{Fast quantum dot single photon source triggered at
  telecommunications wavelength},} Applied Physics Letters \textbf{98}, 083105
  (2011).

\bibitem{Gao:2013df}
W.~Gao, P.~Fallahi, E.~Togan, A.~Delteil, Y.~Chin, J.~Miguel-Sanchez, and
  A.~Imamo{\u{g}}lu, \enquote{Quantum teleportation from a propagating photon
  to a solid-state spin qubit,} Nature Communications \textbf{4}, 2744 (2013).

\bibitem{Rakher:2011ba}
M.~T. Rakher and K.~Srinivasan, \enquote{{Subnanosecond electro-optic
  modulation of triggered single photons from a quantum dot},} Applied Physics
  Letters \textbf{98}, 211103 (2011).

\bibitem{Ates:2013fa}
S.~Ates, I.~Agha, A.~Gulinatti, I.~Rech, A.~Badolato, and K.~Srinivasan,
  \enquote{{Improving the performance of bright quantum dot single photon
  sources using temporal filtering via amplitude modulation},} Scientific
  Reports \textbf{3}, 1397 (2013).

\bibitem{kolchin2008electro}
P.~Kolchin, C.~Belthangady, S.~Du, G.~Yin, and S.~Harris,
  \enquote{Electro-optic modulation of single photons,} Physical Review Letters
  \textbf{101}, 103601 (2008).

\bibitem{Specht:2009cb}
H.~P. Specht, J.~Bochmann, M.~M{\"u}cke, B.~Weber, E.~Figueroa, D.~L. Moehring,
  and G.~Rempe, \enquote{{Phase shaping of single-photon wave packets},} Nature
  Photonics \textbf{3}, 469--472 (2009).

\bibitem{Ma:2014iw}
Y.~Ma, P.~E. Kremer, and B.~D. Gerardot, \enquote{{Efficient photon extraction
  from a quantum dot in a broad-band planar cavity antenna},} Journal of
  Applied Physics \textbf{115}, 023106 (2014).

\bibitem{Flissikowski:2001hx}
T.~Flissikowski, A.~Hundt, M.~Lowisch, M.~Rabe, and F.~Henneberger,
  \enquote{{Photon Beats from a Single Semiconductor Quantum Dot},} Physical
  Review Letters \textbf{86}, 3172--3175 (2001).

\bibitem{Kyoseva:2005ex}
E.~S. Kyoseva and N.~V. Vitanov, \enquote{{Resonant excitation amidst
  dephasing: An exact analytic solution},} Physical Review A \textbf{71},
  054102 (2005).

\bibitem{Ramsay:2010gn}
A.~J. Ramsay, A.~V. Gopal, E.~M. Gauger, A.~Nazir, B.~W. Lovett, A.~M. Fox, and
  M.~S. Skolnick, \enquote{{Damping of Exciton Rabi Rotations by Acoustic
  Phonons in Optically Excited InGaAs/GaAsQuantum Dots},} Physical Review
  Letters \textbf{104}, 017402 (2010).

\bibitem{Malein:2015ur}
R.~N.~E. Malein, T.~S. Santana, J.~M. Zajac, A.~C. Dada, E.~M. Gauger, P.~M.
  Petroff, J.~Y. Lim, J.~D. Song, and B.~D. Gerardot, \enquote{{Screening
  nuclear field fluctuations in quantum dots for indistinguishable photon
  generation},} arXiv:1509.01057v1  (2015).

\bibitem{Varnava:2008bd}
M.~Varnava, D.~E. Browne, and T.~Rudolph, \enquote{{How Good Must Single Photon
  Sources and Detectors Be for Efficient Linear Optical Quantum Computation?}}
  Physical Review Letters \textbf{100}, 060502 (2008).

\bibitem{Jennewein:2011es}
T.~Jennewein, M.~Barbieri, and A.~G. White, \enquote{{Single-photon device
  requirements for operating linear optics quantum computing outside the
  post-selection basis},} Journal of Modern Optics \textbf{58}, 276--287
  (2011).

\bibitem{Santori:2002gg}
C.~Santori, D.~Fattal, J.~Vu{\v{c}}kovi{\'c}, G.~S. Solomon, and Y.~Yamamoto,
  \enquote{Indistinguishable photons from a single-photon device,} Nature
  \textbf{419}, 594--597 (2002).

\bibitem{Laurent:2005cb}
S.~Laurent, S.~Varoutsis, L.~Le~Gratiet, A.~Lema{\^\i}tre, I.~Sagnes,
  F.~Raineri, A.~Levenson, I.~Robert-Philip, and I.~Abram,
  \enquote{{Indistinguishable single photons from a single-quantum dot in a
  two-dimensional photonic crystal cavity},} Applied Physics Letters
  \textbf{87}, 163107 (2005).

\bibitem{Gazzano:2013cq}
O.~Gazzano, S.~Michaelis~de Vasconcellos, C.~Arnold, A.~Nowak, E.~Galopin,
  I.~Sagnes, L.~Lanco, A.~Lema{\^\i}tre, and P.~Senellart, \enquote{{Bright
  solid-state sources of indistinguishable single photons},} Nature
  Communications \textbf{4}, 1425 (2013).

\bibitem{Forstner:2003ku}
J.~F{\"o}rstner, C.~Weber, J.~Danckwerts, and A.~Knorr,
  \enquote{{Phonon-Assisted Damping of Rabi Oscillations in Semiconductor
  Quantum Dots},} Physical Review Letters \textbf{91}, 127401 (2003).

\bibitem{Machnikowski:2004id}
P.~Machnikowski and L.~Jacak, \enquote{{Resonant nature of phonon-induced
  damping of Rabi oscillations in quantum dots},} Physical Review B
  \textbf{69}, 193302 (2004).

\bibitem{RevModPhys.70.1003}
K.~Bergmann, H.~Theuer, and B.~W. Shore, \enquote{Coherent population transfer
  among quantum states of atoms and molecules,} Rev. Mod. Phys. \textbf{70},
  1003--1025 (1998).

\bibitem{PhysRevLett.99.097401}
X.~Xu, Y.~Wu, B.~Sun, Q.~Huang, J.~Cheng, D.~G. Steel, A.~S. Bracker,
  D.~Gammon, C.~Emary, and L.~J. Sham, \enquote{Fast spin state initialization
  in a singly charged inas-gaas quantum dot by optical cooling,} Phys. Rev.
  Lett. \textbf{99}, 097401 (2007).

\bibitem{brunner2009coherent}
D.~Brunner, B.~D. Gerardot, P.~A. Dalgarno, G.~W{\"u}st, K.~Karrai, N.~G.
  Stoltz, P.~M. Petroff, and R.~J. Warburton, \enquote{A coherent single-hole
  spin in a semiconductor,} Science \textbf{325}, 70--72 (2009).

\end{thebibliography}

\begin{thebibliography}{1}
\newcommand{\enquote}[1]{``#1''}

\bibitem{puri2001mathematical}
R.~R. Puri, \emph{\emph{Sections 8.5-7 in} Mathematical methods of quantum
  optics}, vol.~79 (Springer Science \& Business Media, 2001).

\bibitem{breuer2002theory}
H.-P. Breuer and F.~Petruccione, \emph{\emph{Sections 3.3-4 in} The theory of
  open quantum systems} (Oxford university press, 2002).

\bibitem{orszag2007quantum}
M.~Orszag, \emph{Quantum optics: including noise reduction, trapped ions,
  quantum trajectories, and decoherence} (Springer Science \& Business Media,
  2007).

\bibitem{Schaibley:2013bt}
J.~R. Schaibley, A.~P. Burgers, G.~A. McCracken, D.~G. Steel, A.~S. Bracker,
  D.~Gammon, and L.~J. Sham, \enquote{{Direct detection of time-resolved Rabi
  oscillations in a single quantum dot via resonance fluorescence},} Physical
  Review B \textbf{87}, 115311--5 (2013).

\end{thebibliography}



\end{document}